\documentclass[preprint,review,12pt]{elsarticle}

\usepackage{graphicx}
\usepackage[utf8]{inputenc}
\usepackage{amssymb,amsmath}
\usepackage{listings}
\usepackage[nolist,nohyperlinks]{acronym}
\usepackage{color}
\usepackage{sidecap}
\usepackage{verbatim}
\sidecaptionvpos{figure}{c}
\usepackage[hidelinks]{hyperref}

\definecolor{dkgreen}{rgb}{0,0.6,0}
\definecolor{gray}{rgb}{0.5,0.5,0.5}
\definecolor{mauve}{rgb}{0.58,0,0.82}
\graphicspath{{figures/}}

\journal{Computer Physics Communications}

\begin{document}
\newcommand{\somakeywords}{OpenACC, GPU, SCMF, HPC }
\begin{frontmatter}


\title{Multi-Architecture \ac{MC} Simulation of Soft Coarse-Grained Polymeric Materials:\\\acf{SOMA}}
\author[gaug]{L.~Schneider\corref{cor1}}
\ead{ludwig.schneider@theorie.physik.uni-goettingen.de}
\author[gaug]{M.~M\"uller}

\address[gaug]{Georg-August Universit\"at G\"ottingen, Institute for Theoretical Physics, Friedrich-Hund-Platz 1, 37077 G\"ottingen, Germany}
\cortext[cor1]{Corresponding author}

\begin{abstract}

  Multi-component polymer systems are important for the development of new materials because of their ability to phase-separate or self-assemble into nano-structures. The \acf{SCMF} algorithm in conjunction with a soft, coarse-grained polymer model is an established technique to investigate these soft-matter systems. Here we present an implementation of this method: \ac{SOMA}. It is suitable to simulate large system sizes with up to billions of particles, yet versatile enough to study properties of different kinds of molecular architectures and interactions.

  We achieve efficiency of the simulations commissioning accelerators like GPUs on both workstations as well as supercomputers. The implementation remains flexible and maintainable because of the implementation of the scientific programming language enhanced by OpenACC pragmas for the accelerators.

  We present implementation details and features of the program package, investigate the scalability of our implementation \ac{SOMA}, and discuss two applications, which cover system sizes that are
  difficult to reach with other, common particle-based simulation methods.

\end{abstract}


\end{frontmatter}

{\bf PROGRAM SUMMARY}


\begin{small}
\noindent
{\em Program Title:} \acs{SOMA}                                          \\
{\em Licensing provisions:} GNU Lesser General Public License version 3 \\
{\em Programming language:} C99, OpenACC, OpenMP, \acs{MPI}, python  \\
{\em Program obtainable from:}  https://gitlab.com/InnocentBug/SOMA \\
{\em Designed for:} \acs{HPC} clusters and workstations \\
{\em Operating systems or monitors under which the program has been tested:} Linux\\
{\em Has the code been parallelized?: } yes\\
{\em Distribution format:} GIT repository 
{\em Typical running time:}\\
Problem-size dependent. Scales with problem size and the used parallel hardware. Typically less than one day.\\
{\em Nature of problem:}\\
Efficient simulation of polymer materials, their phase-separation or self-assembly using a highly coarse-grained, soft, particle-based model [1]. The simulations help predicting self-assembled structures that, for example, find application in the fabrication of large-scale, dense arrays of nano-structures by \acf{DSA}.\\
{\em Solution method:}\\
Representation of soft, non-bonded interactions by quasi-instantaneous fields on a collocation grid using the \acf{SCMF} [2] algorithm, and sampling of configuration space using  local random \acf{MC} displacements and \acf{SMC}. Parallelization using MPI and accelerators such as \acp{GPU}. \\
{\em Restrictions:}\\
The program has not been tested for more than 10 billion particles. \\
{\em Unusual Features:}\\
Efficient simulation on different hardware architectures and accelerators, including multi-core \acp{CPU} and \acp{GPU}. Furthermore, it is possible to combine different architectures within a single simulation.\\

\end{small}

{\bf LONG WRITE-UP}\\
\section{Introduction}
\label{sec:introduction}

Multi-component polymer melts can exhibit micro- or macrophase separation on scales from nanometers to micrometers. Specifically we consider copolymer melts that are dense liquids of macromolecules, where the flexible, chain-like molecules are composed of two blocks -- $A$ and $B$ -- that are thermodynamically incompatible. These diblock copolymers are characterized by their ability to self-assemble into dense periodic structures on the nanoscale \cite{Leibler80,matsen2001standard} with a correlation length, distance between defects, or grain size that extend to micrometers, millimeters, or even beyond \cite{li2015defects,Li2015pps}. These features qualify them for a wide range of applications ranging from microelectronic device fabrication \cite{Morris15,Stoykovich10,stoykovich2007directed,Stoykovich:2006ha} to filtration membranes \cite{Abetz2014}.

At sufficiently large incompatibility between the constituent blocks, the thermodynamic equilibrium structure is a macroscopically ordered array of domains, in which one of the components is enriched. Depending on the volume fraction of the components and the molecular architecture a wide variety of spatially modulated equilibrium structures \cite{Leibler80} -- e.g., lamellae, cylinders that arrange on an hexagonal lattice, or spherical domains of the minority component that order on a body-centered-cubic \ac{BCC} lattice -- is obtained.

In experiments, however, periodic structures with long-range order are typically not obtained, even if samples are annealed at elevated temperature or in a plasticizing solvent for hours or days \cite{Hur2015}. Instead, the kinetics of structure formation becomes trapped in metastable states that are riddled with defect or consist of multiple grains. Defect annihilation or grain-boundary motion is protracted because it may involve high free-energy barriers \cite{LiMacro2016,Hur2015b,Li2014,Umang12}.

Whereas defect annihilation and grain growth has been studied in great detail in the context of hard-condensed matter systems, such as atomic crystals, there are several, important differences to self-assembled copolymer systems \cite{li2015defects}: (i) In hard-condensed matter, a unit cell typically is composed of only a few atoms, whereas in soft matter a \ac{BCC} sphere contains many macromolecules. (ii) The defect annihilation mechanisms can greatly differ between soft-matter and hard-matter systems. For example, two microphase-separated domains can fuse and thereby reduce the number of unit cells, whereas the number of unit cells is typically conserved in hard crystals because of the atomic nature of their unit cell.

The computational study of structure formation in multi-component polymer melts, however, poses a significant computational challenge for it requires both, (i) large system sizes to simultaneously resolve the properties of the interfaces between $A$ and $B$ domains and the molecular architecture on the nanoscale and the large-scale morphology on the micrometer scale, and (ii) long times to allow for large-scale morphology changes. As a consequence many open questions concerning defects and the kinetics of structure formation in soft-matter systems have remained unanswered.

With the powerful implementation of the \ac{SCMF} algorithm \cite{daoulas2006single} in conjunction with a soft, coarse-grained model \cite{mmlong} we aim to provide a tool to understand the fascinating physics of these complex fluids. \acf{SOMA} is an implementation of the \ac{SCMF} algorithm for a versatile, soft, coarse-grained, particle-based model of multi-component polymer systems with different macromolecular architectures. This specific choice of model and algorithm allows us to optimize aspects of the implementation, which conventional \acf{MD} and \acf{MC} program packages
\cite{hoomd,plimpton2007lammps,berendsen1995gromacs,phillips2005scalable} cannot.

Modern clusters like TITAN of the Oak Ridge Leadership Computing Facility at the Oak Ridge National Laboratory or JURECA \cite{krause2016jureca} at the Neumann Institute for Computing in J\"ulich, Germany, attach to their compute nodes accelerators like Nvidia \acp{GPU}. These accelerators are designed for the parallel execution of the same instruction and their memory is optimized to enable a higher memory throughput than conventional \acp{CPU}, in addition to their better energy efficiency.

Instead of implementing \ac{SOMA} in one specific language for one specific accelerator (e.g., the \ac{CUDA} of Nvidia for Nvidia \acp{GPU}), we decided to implement the acceleration using the OpenACC \cite{openacc} \texttt{\#pragma}-based technique. This enables us to compile a single code base for multiple architectures. While this strategy may not yield the optimal performance compared a more specialized implementation, our approach keeps the code maintainable and flexible even for future accelerators.

The presented \ac{SOMA} software implements the \ac{SCMF} algorithm for multiple accelerated \ac{HPC} clusters, but it is also efficient for workstations if they have an accelerator.

\section{Particle-based model and algorithm}\label{sec:model}

The self-similar structure of long, flexible macromolecules imparts a large degree of universality onto the properties of multi-component, polymer systems. Therefore we use a coarse-grained, particle-based model where each coarse-grained interaction center -- denoted as ``particle'' in the following -- represents a multitude of monomeric repeat units along the backbone of a macromolecule. These coarse-grained segments interact via soft, pairwise potentials \cite{daoulas2006single, mmlong}  that represent the relevant interactions -- connectivity along the molecular backbone, repulsion between between segments -- and that have a computationally convenient form. In our top-down modeling approach the strength of the interactions is related to experimentally accessible observables like the molecules' end-to-end distance, the isothermal incompressibility, or the Flory-Huggins parameter \cite{daoulas2006single, mmlong}. In the following, we focus on the definition of the model and its consequences for our implementation, highlighting its versatility and efficiency.

\subsection{Soft, coarse-grained model}\label{sec:governing-equations}

A macromolecule is defined as a collection of particles that are linked by bonds. We explicitly exclude bonds between different molecules, but otherwise any topology of bonds is permissible. Thus, a macromolecule is defined as bonded network. The system is comprised of $m_t$ distinct types of macromolecular architectures, which differ in their bond topology.

The bonded interactions
\begin{align}
  \label{eq:h-bond}
  \frac{\mathcal{H}_\text{b}}{k_\text{B}T} = \sum_{m\in\{\text{mol\}}} \sum_{b\in\{\text{bonds\}}_m} V_{m,b}(\boldsymbol{r})
\end{align}
are the summed interaction of bonds $b$ in each molecule $m$. The set of bonds $\{ \text{bonds} \}_m$ defines the architecture of the molecule. The interaction energy of each bond $V_{m,b}(\boldsymbol{r})$ can take any non-singular form as a function of the distance, $\boldsymbol{r}$, between the bonded beads. The most common form is a harmonic potential
\begin{align}
  \label{eq:V-harmonic}
  V_{\text{harm.}}(\boldsymbol{r}) = \frac{1}{2} k_{0} \boldsymbol{r}^2
\end{align}
with the spring constant $k_{0}$. This specific form of the bond potential can be justified by the coarse-graining of a Gaussian chain and is a universal feature of highly coarse-grained polymer models. Via this bonded interaction, the energy of a particle directly depends on all its bonded neighbors.

Particles can be of one of $n_t$ different bead types. The bead type dictates the non-bonded, short-range interactions. In our highly coarse-grained, top-down model, these interactions are used to (i) restrain fluctuations of the particle density ($\mathcal{H}_{\text{fluc.}}$) in the nearly incompressible fluid, and (ii) represent the interactions between the $n_t$ different particle types ($\mathcal{H}_{\text{inter.}}$) that give rise to micro- or macrophase separation. The Hamiltonian takes the form $\mathcal{H}_{\text{nb}} = \mathcal{H}_{\text{fluc.}}+\mathcal{H}_{\text{inter.}}$. Inspired by the \ac{SCFT} for multi-component polymer melts, we express these non-bonded interactions as an excess free-energy functional of the local densities, $\hat\phi_i(\boldsymbol{r})$ with $i=0,\cdots,n_t$. The hat indicates that these spatially varying functions depend on the particle positions, $\{\boldsymbol{r}\}$. Formally, the local normalized densities are defined by \cite{mmlong}
\begin{align}
   \hat \phi_i(\boldsymbol{r}) = \frac{1}{\rho_0} \sum_{j\in\{\text{beads}\}}\delta(\boldsymbol{r}-\boldsymbol{r}_j)\delta_{\text{type}(j),i}
\end{align}
where $\rho_{0}$ denotes the average number density of particles in the system. The $\delta$-function is computationally inconvenient, and for a numerical treatment, it can be either mollified by a weighting function, resulting a DPD-like models, or the densities can be evaluated on the grid. The latter scheme is employed in the following, and it is particularly advantageous for dense systems or large invariant degrees of polymerization, when one particle interacts with many neighbors.\footnote{Typically, the range of the interaction, $2\Delta L$, is on the same order as the statistical segment length, $b$, and both microscopic length scales of the model should be small compared to the smallest physical length scale of interest, e.g., the width of the interface between domains. Hence the typical number of particles, with which a reference particle interacts, is given by $8\Delta L^{3}\rho_{0} \approx 8 (\Delta L/b)^{3} \sqrt{\bar{\cal N}/N}$. Using typical values $\bar{\cal N}= 16\,384$ and $N=32$, $\Delta L\approx b$, we estimate that one particle interacts with $181$ neighbors. Evaluating these interacts via a neighbor list is unpractical, and therefore we use a collocation grid to evaluate the non-bonded interactions.}

We use a cubic collocation grid with linear spacing $\Delta L$ and define the densities in grid cell $c$ by the spatial average

\begin{align}
  \label{eq:density-characteristics}
  \hat\phi_i(c) = \frac{1}{\Delta L^{3}}\int \text{d}r\; \Pi_{c}(\boldsymbol{r}) \hat \phi_i(\boldsymbol{r}) = \frac{1}{\Delta L^{3} \rho_0}\sum_{j\in\{\text{beads}\}} \Pi_{c}(\boldsymbol{r}_j) \delta_{\text{type}(j),i},
\end{align}
where the assignment function, $\Pi_c$, is simply taken to be the characteristic function of grid cell, $c$, i.e., if the argument, $\boldsymbol{r}$, is inside the grid cell, $c$, then $\Pi_{c}(\boldsymbol{r})$ is $1$, and $0$ otherwise. The scheme is similar to particle-in-cell models in plasma physics or simple particle-mesh technique to evaluate electrostatic interactions.

Using this collocation grid, we define the non-bonded interactions
\begin{align}
  \label{eq:discrete-ham}
  \frac{\mathcal{H}_\text{nb}[\{\hat\phi_i\}]}{k_\text{B}T} = \frac{\rho_0 \Delta L^3}{N} \sum_{c\in\{\mathrm{cells}\}} \left( \mathcal{K}_{\text{fluc.}}[\{\hat\phi_i(c)\}] + \mathcal{K}_{\text{inter.}}^\alpha[\{\hat\phi_i(c)\}]\right).
\end{align}
Fluctuations of the total density are restrained by the term
\begin{align}
  \label{eq:hamil-comp}
  \mathcal{K}_{\text{fluc.}}[\{\hat\phi_i\}] & = \frac{\kappa_{0} N}{2} \left( \sum_{i=0}^{n_t - 1} \hat\phi_i(c) -1 \right)^2.
\end{align}
where the model parameter, $\kappa_{0}$, is related to the inverse isothermal compressibility. The thermodynamic incompatibility between different particle types, $i\neq j$, is represented by the contribution
\begin{align}
  \label{eq:hamil-pair}
  \mathcal{K}_{\text{inter.}}^0[\{\hat\phi_i\}] &= -\sum_{i\neq j}\frac{\chi_{0ij} N}{4} \big(\hat\phi_{i}(c) -\hat\phi_j(c)\big)^2 \qquad \text{ or}\\
  \mathcal{K}_{\text{inter.}}^1[\{\hat\phi_i\}] &=  \sum_{i\neq j}\chi_{0ij} N \hat\phi_{i}(c) \hat\phi_j(c)
\end{align}
where the model parameters, $\chi_{0ij}$, quantify the thermodynamic incompatibility of different bead types. The first option $\mathcal{H}_{\text{inter.}}^0$ can be found in Ref.~\cite{daoulas2006single}, the latter $\mathcal{H}_{\text{inter.}}^1$ in Ref.~\cite{pike2009theoretically}. Since the grid-based densities are functions of the particle coordinates, the non-bonded interactions, $\mathcal{H}_{\text{nb}}$ depend on $\{\boldsymbol{r}\}$ and are suitable for molecular simulations. Note that quadratic terms in the excess free energy can be explicitly rewritten in terms of pairwise potentials.

$N$ is a reference number of particles per macromolecule. It does not influence the interactions but for systems of linear polymer chains, the properties are invariant under changing the number of particles (discretization), $N$, along the macromolecular contour provided that the interaction strengths, $\chi_{ij}N$ and $\kappa N$, per polymer molecule remain fixed. This quantity is also employed to relate the particle number density, $\rho_{0}$, to the invariant degree of polymerization,

\begin{align}
\bar{\cal N} = \left(\frac{\rho_{0}}{N}R_{e0}^{3}\right)^{2} \approx \left(\rho_{0}b_{0}^{3}\right)^{2}N
\end{align}
where $R_{e0}$ is the root mean-squared end-to-end distance of a polymer chain in the absence of non-bonded interactions, and $b_{0}= \sqrt{3/k_{0}}$ is the statistical segment length of an ideal chain.

Our soft, coarse-grained model is particular suitable for investigating the thermodynamics and kinetics of structure formation of dense multi-component polymer systems \cite{Muller08b}. The softness is very instrumental in modeling systems with in experimentally relevant, large invariant degree of polymerization, $\bar{\cal N}$, where one chain molecule interacts with a multitude of neighbors. Rather than increasing the number of coarse-grained interaction centers along the backbone of the long, flexible macromolecule, we achieve large values of $\bar{\cal N}$ by increasing the particle density at $N \ll \bar{\cal N}$.

\subsection{Sampling algorithm: \ac{SCMF} Monte-Carlo simulation}
\label{sec:sampling-algorithm}
Whereas the thermodynamic properties of our soft, coarse-grained model can be straightforwardly studied by Monte-Carlo simulations, we additionally exploit the difference between the strong but computationally inexpensive bonded interactions, $\mathcal{H}_\text{b}$, and the significantly weaker but computationally costly non-bonded interactions, $\mathcal{H}_\text{nb}$. The difference in interaction strength can be quantified by the scale of the typical forces on a particle
\begin{align}
\boldsymbol{F}_\text{b} &\sim \frac{k_\text{B}T}{b_{0}} \sim \frac{k_\text{B}T}{R_{e0}} \sqrt{N} \\
\boldsymbol{F}_\text{nb} &\sim \frac{\chi_{0} k_\text{B}T}{w} \sim \frac{k_\text{B}T}{R_{e0}} \frac{(\chi_{0}N)^{3/2}}{N}
\end{align}
where $\chi_{0}k_\text{B}T$ denotes the energy of transferring a particle across a domain interface and $w \sim R_{e0}/\sqrt{\chi_{0}N}$ is the interface width, i.e., the typical ratio of bonded to non-bonded force scales increases like $N^{3/2}$ with the chain-contour discretization at fixed $\chi_{0}N$ \cite{mm11b}.

In \ac{MD} simulations such a scale separation is routinely exploited by multiple-time-step integrators like RESPA \cite{tuckerman1992reversible}, and the \ac{SCMF} algorithm can be conceived as an \ac{MC} analog. Thus, in \ac{SCMF} simulations, the slowly varying but computationally costly non-bonded pair interactions are approximated by quasi-instantaneous external fields, $\omega_k(c)$, for the different species, $k$, in each grid cell, $c$. These dimensionless interaction fields (in units of the thermal energy scale, $k_\text{B}T$) are computed from the instantaneous densities according to

\begin{align}
  \label{eq:omega-fields}
  \omega_i(c) &= \frac{1}{k_\text{B}T \rho_0 \Delta L^3}\frac{\partial \mathcal{H}_\text{nb}}{\partial \hat \phi_k(c)} \nonumber\\
  &\overset{\mathcal{K}_\text{inter}^0}{=} \kappa_{0} \left(\sum_{k=0}^{n_t-1} \hat \phi_k(c) -1\right)- \sum_{k\neq i}^{n_t} \frac{\chi_{0i,k}}{2}(\hat\phi_i(c) - \hat\phi_k(c))\nonumber\\
  &\overset{\mathcal{K}_\text{inter}^1}{=} \kappa_{0} \left(\sum_{k=0}^{n_t-1} \hat \phi_k(c) -1\right) + \sum_{k\neq i}^{n_t} \chi_{0i,k} \hat\phi_k(c)
\end{align}
for the respective options of the Hamiltonian.

Thus, a \ac{SCMF} cycle is comprised of two stages \cite{daoulas2006single}:
\begin{enumerate}
\item Computation of the interaction fields, $\omega_{i}$, from the instantaneous particle configuration $\{\boldsymbol{r}\}$ according to Eqs.~(\ref{eq:density-characteristics}) and (\ref{eq:omega-fields}).
\item While keeping $\omega_{i}(c)$ constant, update the particle configuration by a \ac{MC} scheme using the bonded interaction, ${\cal H}_\text{b}$ and the \ac{SCMF} interaction with the external field
\begin{align}
\frac{{\cal H}^\text{SCMF}_\text{nb}}{k_\text{B}T} = \sum_{i=0}^{n_{t}-1}\sum_{c\in\{\mathrm{cells}\}} \Delta L^{3}\rho_{0}\,\hat \phi_{i}(c) \omega_{i}(c) =\sum_{j\in\{\text{beads}\}} \omega_{\text{type}(j)}(c_{j})
\end{align}
where $c_{j}$ denotes the cell, in which particle $j$ is located.
\end{enumerate}

These quasi-instantaneous, non-bonded interactions are an accurate representation of the pairwise interaction of a particle with its instantaneous, fluctuating environment, if (i) $\omega_{i}(c)$ are frequently updated and (ii) the parameter \cite{daoulas2006single}
\begin{align}
\varepsilon = \frac{1}{\sqrt{N\bar{\cal N}}}\left(\frac{b_{0}}{\Delta L} \right)^{3}
\end{align}
is small. For typical values $\bar{\cal N}= 16\,384$, $\Delta L\approx b$, and not too small discretization of the molecular contour, $N=32$, we obtain $\varepsilon\approx 10^{-3}$, and quantitative comparison with \ac{MC} simulations using the exact, non-bonded Hamiltonian, Eq.~(\ref{eq:discrete-ham}), have confirmed that the \ac{SCMF} algorithm can accurately capture the thermodynamic behavior including fluctuations \cite{daoulas2006single}.

In order to study the kinetics of structure formation, the \ac{MC} update in the second stage of the \ac{SCMF} algorithm should approximate the single-chain dynamics and give rise to a locally conserved, diffusive behavior of the densities. Within the \ac{SOMA} program, we have implemented two types of \ac{MC} moves of the particles: (i) local random displacements of a particle accepted by the Metropolis acceptance criterion \cite{metropolis1949monte}, where the trial displacement is uniformly chosen with a maximal step length, $a$, along each Cartesian direction, or (i) a \acf{SMC} scheme \cite{pangali1978novel,rossky1978brownian} that employs the strong bonded forces to propose a trial displacement similar to Brownian dynamics and results in Rouse-like dynamics \cite{MullerSL}. In the parallel implementation we present here (cf.~Sec.~\ref{sec:single-rank-parall}), the detailed-balance criterion of each individual moves ensures global-balance of the concomitant master equation for a parallel step.

Finally, let us highlight the three main advantages of the soft, coarse-grained model in conjunction with the \ac{SCMF} algorithm for studying the thermodynamics and the kinetics of structure formation of multi-component polymer systems
\begin{enumerate}
\item The softness of the interactions allows us to represent polymer systems with a large invariant degree of polymerization, ${\cal \bar N}$, by using a large number density $\rho_{0}$ of particles rather than by increasing the chain discretization, $N$.
\item The calculation of the non-bonded interactions on a collocation grid rather than by  neighbor list is computationally efficient because, for typical parameters, a single particle interacts with $10^{2}$ neighbors whereas the change of the non-bonded energy in response to a particle displacement can simply be computed by
\begin{align}
  \label{eq:delta-e}
  \Delta E_{\text{nb}} = \omega_k(c') - \omega_k(c).
\end{align}
where $c$ and $c'$ denote the grid cells, in which the particle is located before and after the move, respectively.
\item The use of an external field during the update of the molecular conformation in the second stage of the \ac{SCMF} cycle, temporarily decouples different molecules. Thus, the \ac{SCMF} algorithm intrinsically incorporates parallelism between all molecules and allows for an efficient parallelization of the program.
\end{enumerate}

\section{Implementation}\label{sec:implementation}

The general goal of the implementation of \ac{SOMA} is, on the one hand, to obtain a high-performance code for modern computer architectures while, on the other hand, to allow for easy use and modification by the scientific user. For that reason we decided to implement \ac{SOMA} in C
as a common scientific programming language.\footnote{The C99 standard to utilize fixed width data types.} The \texttt{\#pragma}-based acceleration approach of OpenACC \cite{openacc} does not reduce readability, but allows parallel implementations for multiple architectures.

OpenACC supports execution on external accelerators. These accelerators do not necessarily share the memory with the \ac{CPU}. So the role of OpenACC is two-fold: It organizes the memory transfer between the \ac{CPU} and the accelerator, and it generates the parallel code for the accelerator.

Our general aim is to execute all computation on the accelerator while minimizing the interference of the hosting \ac{CPU}. To this end, after initialization, the simulation data is copied  to the accelerator and all computation for the simulation is executed on the accelerator without further memory transfer. Only infrequent calculations of physical observables for on-the-fly analysis of the configuration or \ac{MPI} communications may require updates of the \ac{CPU} memory.

\subsection{Hybrid parallelism}\label{sec:hybrid-parallelism}

For efficient utilization of modern high-performance computers parallelism is important. We employ a hybrid strategy: (i) To connect multiple shared-memory systems, an \ac{MPI}-based parallelization is used. (ii) On shared memory systems, like multi-core \acp{CPU} or an accelerator, the \texttt{\#pragma}-based approaches of OpenMP or OpenACC are used to implement parallel code.

\subsubsection{\ac{MPI}-based parallelization and load-balancing}\label{sec:MPI-level}

The SCMF model introduces an implicit parallelism because all macromolecules interact only via quasi-instantaneous, fluctuating interaction fields. By definition of a  macromolecule, distinct macromolecules are not connected by bonds. At initialization, entire macromolecules are efficiently distributed among the \ac{MPI} ranks via parallel \ac{IO}.

In the first stage of the \ac{SCMF} simulation cycle, each \ac{MPI} rank, $\alpha$, counts the number of particles, $n^{\alpha}_{i}(c)$, of the different segment types, $i$, that stem from its ``own'' molecules, and an \ac{MPI}-allreduce summation combines these grid-based partial occupation numbers of the \ac{MPI} ranks to the density fields, $n_{i}(c) = \sum_{\alpha}n^{\alpha}_{i}(c)=\rho_{0}\Delta L^{3}\hat \rho_{i}(c)$, after each \ac{MC} step. The resulting density fields on the entire grid are available on each \ac{MPI} rank, and each \ac{MPI} rank locally computes the interaction fields, $\omega_{i}(c)$, according to Eq.~(\ref{eq:omega-fields}). We observed that the computation of the interaction fields, $\omega_{i}(c)$, from the occupation numbers, $n_{i}(c)$, can be computed faster on each \ac{MPI} rank than a distributed computation and subsequent communication. The calculation is a simple iteration of grid cells with only cache/coalescence optimized accesses to the memory.

For typical parameters, the number, $n_{i}(c)$, of particles of type $i$ in a grid cell is less or on the order of $\rho_{0}\Delta L^{3}\approx \sqrt{\bar{\cal N}/N}(\Delta L/b)^{3} \sim 10^{2}$. Therefore we safely encode this data as unsigned integers of $16$-bits width. Thus the occupation data of a system with linear dimension $L$ require a total memory of $2 n_{t} (L/\Delta L)^{3} \approx 2 n_{t} N^{3/2} (L/R_{e0})^{3}$ bytes. Compared to the configuration data -- i.e., the spatial positions, $\{\boldsymbol{r}\}$, of all particles require about $24 N \sqrt{\bar{\cal N}} (L/R_{e0})^{3}$ bytes  -- the occupation data, $n_{i}(c)$, is smaller by a factor $\sqrt{\bar{\cal N}/N}(12/n_{t})\sim 10^{2}$. The \ac{MPI} ranks exchange only the occupation data via the \ac{MPI}-allreduce collective sum operation, limiting the required throughput. Depending on the accelerator of the ranks and the \ac{MPI} implementation, this operation might not require an intermediate copy of the occupation data to \ac{CPU} memory. A CUDA-aware \ac{MPI} implementation in combination with Nvidia \acp{GPU} as accelerators enables \ac{MPI} operations directly on the \ac{GPU} memory. This approach scales the application over multiple compute nodes using only a single \ac{MPI}-allreduce operation.

In the second stage of the \ac{SCMF} simulation cycle, each \ac{MPI} rank propagates only its ``own'' molecules via a short \ac{MC} simulation. This part is most compute-intense but does not require \ac{MPI} communication.



Our molecule-based parallelization strategy without spatial domain decomposition enables an easy load balancing between \ac{MPI} ranks: The slowest rank sends a number of its molecules to the fastest rank until the execution time per \ac{MC} step is balanced across all \ac{MPI} ranks. This might be necessary, because the \ac{MPI} ranks might be accelerated by heterogeneous accelerators, or the size of molecules differ among ranks.

\subsubsection{Single-rank parallelism}\label{sec:single-rank-parall}

The second level of our hybrid strategy concerns the parallelism in each \ac{MPI}-rank on a shared-memory system or the use of an accelerator. In the first stage of the \ac{SCMF} simulation cycle, the occupation numbers, $n_{i}(c)$, are gathered and the interaction fields, $\omega_i(c)$, are computed, whereas in the second stage the particle configuration is propagated by a short \ac{MC} simulation.

The computational effort of both stages scales with the number of particles, however, profiling showed that on all tested computer architectures the first stage only takes a fraction the computation time compared to the second one. One reason is that the \ac{MC} moves requires multiple expensive generations of pseudo random numbers. As a consequence, we focus in the following only on the \ac{MC} moves.

In the course of a \ac{MC} move, we choose a sequence of particles independent from the instantaneous configuration, $\{\boldsymbol{r}\}$, try to displace each particle using local, random displacement or \ac{SMC} proposals, and accept each attempt according to the Metropolis acceptance rate. In the second stage of the \ac{SCMF} cycle, particles interact via the bonded interactions and with the temporarily constant, interaction fields, $\omega_i(c)$. Thus, particles that are not directly connected by bonds to each other can be moved independently and simultaneously because the outcome of the \ac{MC} lottery of one particle does not depend on or influence the outcome of the \ac{MC} lottery of the other particle. In order to exploit this intrinsic parallelism, we use a hierarchical scheme:

\begin{itemize}

\item{\em 1. Parallelization over distinct molecules:}\label{sec:polymer-iteration}
By definition, particles that belong to distinct molecules are not connected by bonds and therefore can be moved independently and simultaneously.  Thus, we assign each molecule to a parallel thread, i.e., each thread processes an entire molecule by sequentially choosing a particle of this molecule at random for a \ac{MC} move. On average, each particle of the molecules is attempted to be moved once during a \ac{MC} step. This simple parallelization over distinct molecules is efficient, provided that each \ac{MPI} rank owns sufficiently more molecules than the maximal number of parallel threads. Conventional multi-core systems fulfill this requirement, but highly parallel accelerators may be undersaturated with systems of moderate sizes. \autoref{code:polymer} shows the implementation of this scheme using OpenMP and OpenACC.

\begin{lstlisting}[captionpos=b,caption=Parallel Molecule Iteration Scheme: implemented with OpenMP for shared memory systems and OpenACC for accelerators.,label=code:polymer]
//parallel iteration of all molecules on rank
#pragma acc parallel loop vector_length(tuning_parameter)
#pragma omp parallel for
for(uint64_t mol=0; mol < N_local_mol; mol++){
  const unsigned int N = get_N(mol);
  // sequential iteration of particles
#pragma acc loop seq
  for(int mono=0; mono < N; mono++ ){
    //select random particle to move
    const unsigned int i = rng(N);
    //Monte-Carlo scheme for particle i
    }
  }
\end{lstlisting}

\item{\em 2. Iteration over independent particles within a molecule / ``SET'' algorithm:}\label{sec:iter-indep-sets}

\begin{figure}
  \centering
  \includegraphics[width=\columnwidth]{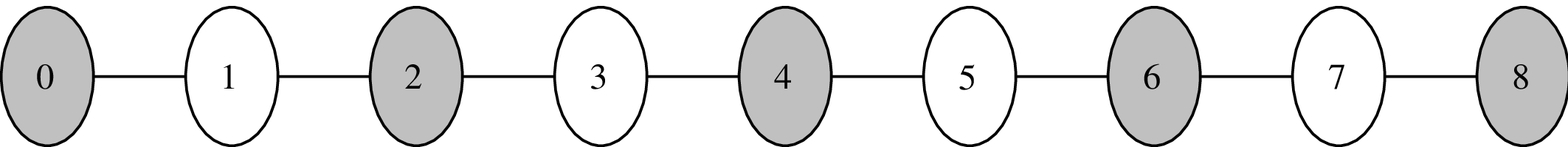}
  \caption{Example of a linear polymer molecule with $N=32$ monomers. Since particles are only bonded to their two nearest neighbors, every second monomer is independent. The coloring marks the two independent sets.}
  \label{fig:sets}
\end{figure}

The second approach exploits a  finer level of parallelism by identifying sets of independent particles within a single molecule. All particles within one set belong to the same molecule but are not directly bonded with one another and therefore can be moved independently and simultaneously. An example for such a partitioning for a linear molecule is depicted in \autoref{fig:sets}. All even particles belong to set $0$, whereas all odd particles belong to set $1$. Our implementation of set generation uses a simple heuristics for the initialization of the sets of independent particles, aiming for the minimal number of sets with a large and approximately equal number of independent particles in each set. Since the architecture of a molecule type is fixed in the course of the simulation, sets of independent particles for each type of molecules only need to be computed at initialization and remain constant afterwards.

A full \ac{MC} move sweeps in parallel over all distinct molecules like in previous scheme. For each molecule, the algorithm sequentially works on every set in a pre-defined random order. The independent particles that belong to the same set are assigned to parallel threads that attempt the \ac{MC} move. Thus the total number of parallel threads is the product of the number of polymers and the (minimal) number of independent particles in a set. \autoref{code:sets} demonstrates the implementation of this second layer of parallelism in OpenACC.

\begin{lstlisting}[captionpos=b,caption=\ac{MC} scheme iteration with two levels of parallelisation utilizing independent sets.,label=code:sets]
#pragma acc parallel loop vector_length(tuning_parameter)
#pragma omp parallel for
for (uint64_t mol = 0; mol < N_local_mol; mol++){
  // Generate random permutation of the sets
  // and store result in set_permutation[]

  // Flat 2d array, sorting elements in independent sets.
  // Computed at initialization and only accessed now.
  const unsigned int* sets = get_sets(mol);

  //Iterate sets in random order.
#pragma acc loop seq
  for(unsigned int i_set=0; i_set < n_sets; i_set++){
    const unsigned int set_id = set_permutation[i_set];
    // Second parallel layer iteration of set elements
#pragma acc loop vector
    for(unsigned int i_p=0; i_p < set_length[set_id]; i_p++){
      unsigned int ip = sets[ set_id*max_member + i_p];
      //MC scheme for particle ip
      }
    }
  }
\end{lstlisting}

In this scheme, the attempt probability for each particle is exactly $1$ as opposed to the former scheme, yielding a slightly faster dynamics. We also note that the sequential iteration over sets breaks detailed balance but global balance is still obeyed because the sequence of particle moves is independent from the configuration, $\{\boldsymbol{r}\}$.

\end{itemize}

\subsection{\ac{GPU} Optimization}\label{sec:gpu-optimization}

The most powerful accelerator we have access to at high-performance computing centers are Nvidia \acp{GPU} (K20, K80, and P100). The main optimizations discussed in this section are tailored to these devices but most of them are also beneficial or almost neutral for the \ac{CPU} performance.

The \ac{GPU}-computation model differs from the \ac{CPU} model: Because of the large number of parallel threads and the partly single-instruction-multiple-data restrictions, actual computations are
rather inexpensive on a \ac{GPU}. On the other hand, however, memory throughput is more important. Thus, memory-access patterns have to be optimized, and they may differ from \ac{CPU}-cache optimized
patterns. Moreover, communication between threads is complicated on a \ac{GPU}. The latter aspect, however, is less significant because our \ac{SCMF} algorithm does not require communication in the course of the computationally expensive \ac{MC} propagation (second stage of \ac{SCMF} cycle).

\subsubsection{Compression of molecular architecture}\label{sec:compr-molec-arch}

Since memory throughput and memory-access pattern are crucial for performance on \acp{GPU}, we optimized the way information is stored. For a \ac{MC} move each particle needs to know its own position, type, interaction field, and the neighbors, to which it is directly bonded, as well as the concomitant bond types.

Typically, the system is comprised of many molecules with same molecular architecture. Rather than storing the properties, type and bond information, for each particle in the molecule separately, we store this information globally for all molecule types and assign each molecule a type. Each particle can infer from the molecule type and its ID in the molecule, to which molecular architecture it belongs to and all other required information.

In the best case, every molecule is of the same type and the architecture information is only stored once for all molecules. In the worst case, every molecule is of a different type, thus every molecule needs its own architecture memory block. However, this is identical to the amount of memory that would be be required if the architecture information were stored for every molecule individually. Therefore, for the most common scenarios, the amount of stored memory and therefore its throughput is significantly decreased. Moreover, the memory-access pattern is optimized as well: Consider the parallelization over distinct molecules, where each thread in a \ac{GPU} warp operates on a molecule. By initialization we can ensure that neighboring threads operate on molecules of the same type and, therefore, neighboring threads read the same memory area containing the information of the architecture of this molecule type. In the case of iteration over sets of independent particles within a molecule, the threads of a warp are operating on particles that belong to a single molecule and, consequently, access the same memory area  containing the architecture information. Thus, by the compression of architecture for multiple molecules of the same type, we can reduce the required memory throughput as well as we create collaborative memory patterns.

\begin{figure}
  \centering
  \includegraphics[width=\columnwidth]{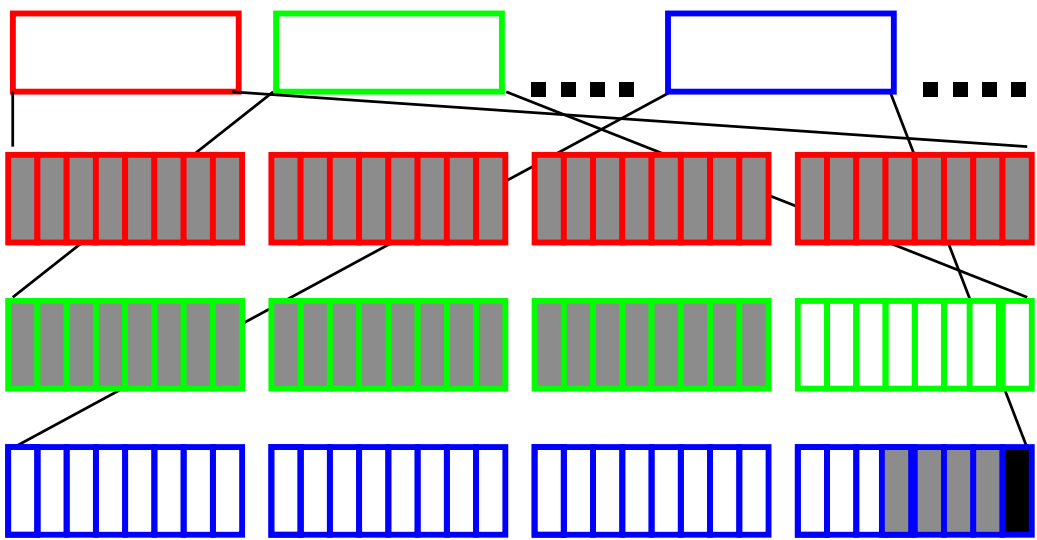}
  \caption{Layout of the memory array that contains the molecular architecture information: Each molecule starts with a single $4$ byte element (red) quantifying the number, $N$, of particles (monomers) in this molecule by an unsigned integer. This length information is followed by $N$ elements (green) of $4$ bytes that contain two pieces of information about each monomer with ID$=0,\cdots,N-1$: The first 3 bytes (gray) define an offset, where the bond list information for this monomer starts in the array. The last byte (white) defines the type of the particular monomer. This monomer array is followed by a list of bond information (blue) that enumerates the bonds between monomers inside the molecule. Each bond is represented by $4$ bytes -- the first $27$ bits (white) represent a signed integer, which defines the offset between the current monomer ID and its bonded partner. The subsequent $4$ bits (gray) identify the type of the bond. Currently only harmonic bonds are implemented, but the concept allows for an easy extension. The last bit (black) signals the last bonded partner of the monomer. The list elements are meant to be iterated; the iteration stops as soon as the last bit (black) is set to $1$. Thus, each molecule can be bonded to an arbitrary number of neighbors.}
  \label{fig:polyarch}
\end{figure}

Technically, the information of the molecular architecture of all types is stored a single array of $32$ bit memory pieces. \autoref{fig:polyarch} describes the memory layout in detail. This approach allows an easy handling of the information and is making sure that the access pattern of the threads is condensed to this single array.

\subsubsection{Autotuner}\label{sec:autotuner}

Modern accelerated environments provide a hierarchical parallelism on a single accelerator. The OpenACC framework offers three layers of parallelism: gang, worker, and vector. The performance of an application may depend on the distributions of the work among the layers, and the optimal distribution, in turn, is influenced by the parameters of the system and algorithm. For example, the simulation that uses the algorithm of partitioning molecules into sets of independent particles will often work best, if the lowest parallel level executes exactly one molecule.

As a consequence, we automatically adjust the distribution at run time. Every critical parallel section is tuned by a separate Autotuner. The Autotuner tests all available parallel configurations and measures the computation time. After an equilibration phase, the fastest configuration is chosen for the rest of the simulation. This approach is similar to the Autotuner approach of the HOOMD-blue\cite{hoomd} simulation package in the CUDA framework.

\subsubsection{Pseudo-Random-Number Generation}\label{sec:pseudo-random-number}

For \ac{MC} simulations \ac{PRNG} is crucial. The quality of the numbers needs to be sufficient to guarantee correctness of the algorithm and computed physical properties and, additionally, speed is a critical factor. There are multiple approaches to utilize \ac{PRNG} in highly parallel environments. The complexity arises because each thread needs random numbers of high quality with no correlation to its fellow threads.
There are three approaches: First, a single thread generates with conventional techniques random numbers and distributes them across all threads. This approach, however, is not practical for highly parallel environments because of its serial nature and the limited efficiency of communication between threads. The other approaches require a separate internal state of the generator for each thread. This state can either be hashed from the thread ID and the time step, at every time it is needed \cite{hoomdRN}, or initialized once and stored in the device memory, where each thread loads and stores its individual state for the generation of random numbers.

We selected the latter strategy because this scheme is efficient on both architectures, \ac{CPU} and \ac{GPU}, for small internal states.
In the case of multiple \ac{PRNG} in a single thread, the internal state is first loaded into register memory, where all generation takes place, and afterwards stored back. For the \ac{GPU} efficiency, a small internal state is crucial because the algorithm is memory bound.

We have selected the \ac{PCG}\cite{pcg} random number generator with an internal state of $128$ bits and stream capability. For comparison, the internal state of the widely used \ac{MT} \cite{matsumoto1998mersenne} is about $150$ times larger. Although the maximal performance is achieved with the \ac{PCG}, we additionally implemented two alternative \acp{PRNG}: the \ac{MT} and a variant with a smaller internal state, the
TT800 \cite{matsumoto1994twisted}. The alternative \ac{PRNG} can be selected for verification of results.

\section{Integrated features}\label{sec:features}
In this section, we briefly summarize the most useful features of the scientific work flow integrated into the \ac{SOMA} simulation package.

Our implementation, using a hybrid parallelism with \ac{MPI} and OpenACC/OpenMP, supports a variety of hardware environments. The scientific user can choose between two different schemes for selecting particles for parallel configuration updates and two different \ac{MC} algorithms -- random, local displacements and \ac{SMC} moves. For highly connected molecules, e.g., star polymers or dendrimers, the equilibration of the internal molecule and its interaction with the environment can be decoupled, by applying simple \ac{MC} moves to the center of mass of the molecule (albeit the quasi-instantaneous field approximation of the \ac{SCMF} algorithm becomes significantly less accurate). The package implements three different \acp{PRNG} -- \ac{PCG}, \acf{MT}, and TT800 -- and supports floating point operations in single and double precision.

Furthermore, we use the \ac{HDF5} \cite{hdf5} to store platform independent all simulation data in a binary format. By employing the \ac{MPI} parallel features of the \ac{HDF5}, we distribute all \ac{IO} across the different \ac{MPI} ranks; each rank touching only the data it requires. In addition, we offer conversion tools to visualize simulation data with ParaView\cite{paraview} using the \ac{XDMF}.

To adjust the simulation to different physical situations we provide an \ac{XML} input format, which is later converted to \ac{HDF5}. The \ac{XML} input file allows the specification of two different Hamiltonians $\mathcal{H}$ (cf.~\autoref{eq:hamil-pair}), and the molecular architecture can be described in a variation of the CurlySMILES \cite{drefahl2011curlysmiles} standard, which allows almost arbitrary architectures with up to $255$ different particle types. Any mixture of different molecular architectures is possible.

We further implemented support for two different kinds of external fields. The first prevents any particle from entering a specific grid cell, enabling simulations of geometric confinements. The second specifies a linear attraction for the particle of a specific type. For instance, this feature allows for the study of wetting phenomena or permits the modeling of guiding structures in \ac{DSA} studies.

\section{Results and discussion}\label{sec:results}
In this section we report for the computational performance of the \ac{SOMA} program package. We compare different hardware configurations including several generations of Nvidia \acp{GPU} and \acp{CPU} as well as different system sizes and configurations. Finally, we demonstrate that we are able to use different architectures during a single simulation run, enabling the efficient use of modern super computers.

\subsection{Strong scaling}\label{sec:strong-scaling}
Strong scaling demonstrates the scalability of code to multiple compute nodes, i.e., at fixed system size we increase the number of processes. Additionally, we use this test for a first comparison of different architectures and algorithms. The code is tested on JURECA \cite{krause2016jureca} at the Neumann Institute for Computing in J\"ulich, Germany, where each node has two K80 \acp{GPU} (four visible devices) and two Intel Xeon E5-2680v3 Haswell \acp{CPU}.

For \ac{CPU} nodes we place a single \ac{MPI} rank on each node and parallelize on each node using the shared memory parallelism of OpenMP. This minimizes the number of \ac{MPI} ranks and therefore optimizes the \ac{MPI} communication patterns. For \ac{GPU} accelerated runs, each accelerator is operated by its own \ac{MPI} rank.

As a reference system we use a melt of linear homopolymers with $\sqrt{\bar{\cal N}}\approx 156$ comprised of $N=64$ beads each and $\Delta L\approx 0.17 R_{e0}$. To cover the scale of prospective simulations, we use two different sizes with a total of $nN=10^7$ and $nN=10^8$ particles. The performance is quantified by the \ac{MPTPS} unit.

\begin{figure}
  \centering
  \includegraphics[width=\columnwidth]{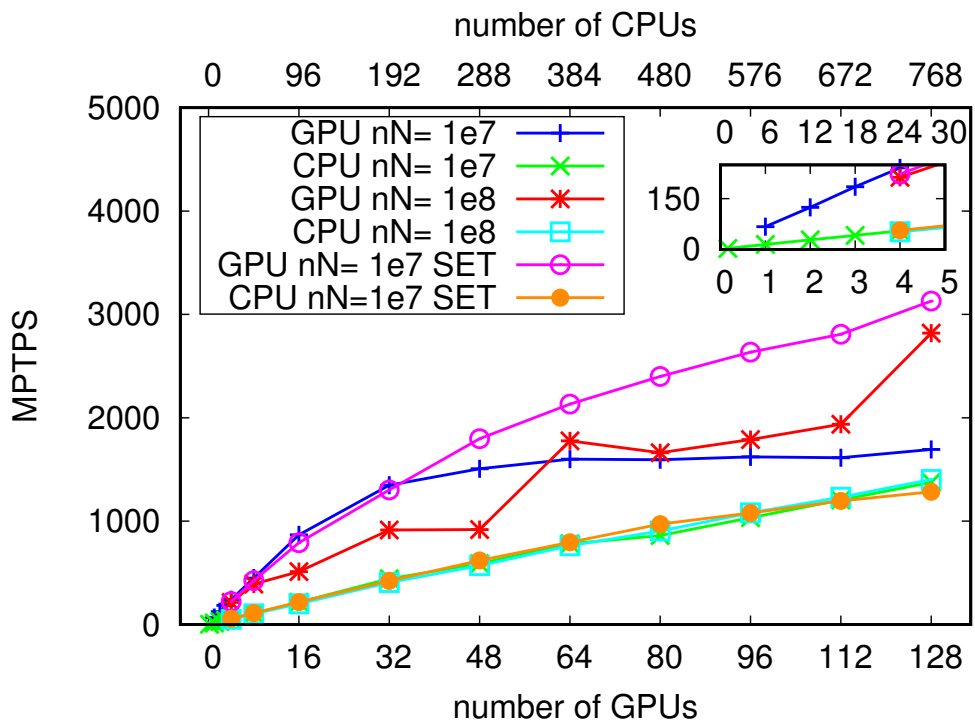}
  \caption{Scaling of \ac{SOMA} for two different system sizes: $nN=10^7$ and $nN=10^8$ particles, on the JURECA system. The performance is quantified by Million Particle Timesteps Per Second (MPTPS). ``SET''-marked results were obtained by iterating over sets of independent particles within a molecule. The inset demonstrates the good linear scaling on the level of a single node.}
  \label{fig:soma-scaling}
\end{figure}

The strong scaling of program package is presented in \autoref{fig:soma-scaling}. 
We find that the \ac{CPU} performance of \ac{SOMA} scales almost perfectly linearly. Nevertheless, the \ac{GPU} performance outspeeds the \ac{CPU} implementation for every size and node configuration. Thus, it is always beneficial to use the \ac{GPU} implementation, if available.

The scaling for \acp{GPU}, however, is not as good as for the \ac{CPU}, and there are two aspects explaining this effect:
(i) The smaller system, $N=10^7$, does not scale for configurations with more than $32$ \acp{GPU} because the parallelization over distinct molecules is not able to saturate the \ac{GPU} anymore. In that configuration, each \ac{GPU} owns only approx.~$4882$ polymers and therefore  each \ac{GPU} has only about $ 4882$ threads. This is clearly not enough for an Nvidia K80 to hide all memory latencies. This performance issue can be mitigated by the iteration over sets of independent particles (``SET'' algorithm) where a total of $nN/2$ threads are available instead of only $nN/64$. Thus the \acp{GPU} can be much better saturated and the performance almost doubles for $nN=10^{7}$ and $128$ \ac{GPU}.
(ii) For the bigger system, $nN=10^8$, this problem does not arise anymore, but the \ac{MPI} all-to-all communication of the array of occupation numbers, $n_i(c)$, now limits the performance. This is easily identified because for a number of ranks of a power of $2$, which is optimal for binary-tree communication patterns, the performance is significantly better.

\subsection{Weak scaling}\label{sec:weak-scaling}

\begin{figure}
  \centering
  \includegraphics[width=\columnwidth]{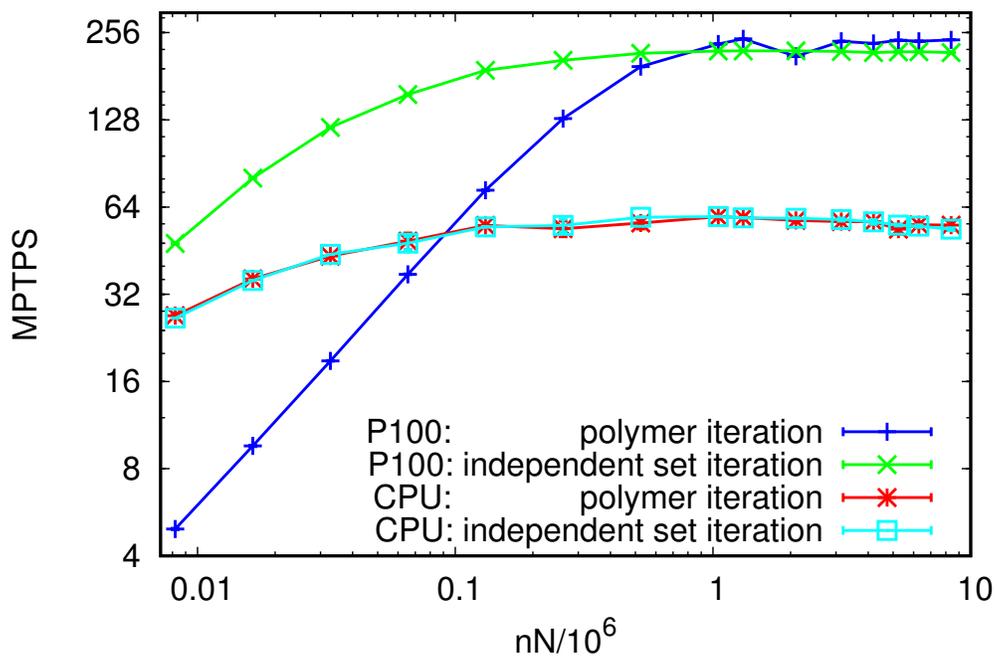}
  \caption{Weak scaling for a homopolymer melt with $N=64$ particles per molecule. The better efficiency of the ``SET'' algorithm on a \ac{GPU} architecture for small system sizes becomes clear.}
  \label{fig:weak}
\end{figure}

To further investigate the two iteration algorithm and their effect on performance on different architectures, we perform a weak-scaling analysis. To this end we simulate a simple homopolymer melt
with $N=64$ particles per molecule and vary the number of molecules $n\in[2^7:2^{17}]$, while maintaining a constant $\sqrt{\bar{\cal N}} \approx 128$ and $\Delta L \approx 1/6 R_{e0}$. Two different hardware configurations are compared: one Pascal P100 Nvidia \ac{GPU} and a JURECA node with $24$ \ac{CPU} cores. \autoref{fig:weak} graphically presents the achieved performance. For simulations using accelerators and parallelization over molecules, the performance dramatically decreases as we decrease the system size below $nN=10^{6}$. The reason is, as noted before, the insufficient saturation of the \ac{GPU} devices. The higher parallelism on the level of independent particles instead of molecules for the ``SET'' algorithm significantly reduces the deterioration of the performance for the small system sizes. For the smallest system, $n=2^{7}$, i.e., $nN=8192$ particles, the ``SET'' algorithm is more than an order of magnitude more efficient than the parallelization over molecules.

For larger systems the performance stabilizes to a plateau, which is the best performance of a saturated device. In this case the iteration over independent molecules is slightly more efficient than the ``SET'' algorithm. The reason for this characteristic is the downside of the higher parallelism: Each thread requires an internal state for the \ac{PRNG}. As a consequence, the ``SET'' algorithm requires an higher memory throughput per iteration step, resulting in a slight slowing down of the memory-bound algorithm. In the present scenario, the ``SET'' algorithm allocated about $27\%$ more device memory than the molecule-parallel counterpart.

The \ac{CPU} performance is unaffected by the choice of the iteration algorithm because it cannot benefit of the higher degree of parallelism. On the other hand, however, it also does not suffer as much from the higher memory throughput. For small systems, which are not using the GPU optimized ``SET'' algorithm, it can be beneficial to run on a multicore machine compared to an accelerator. But using either the ``SET'' algorithm or larger systems, the P100 accelerator outspeeds the $24$ \ac{CPU} cores by roughly a factor of four.

\subsection{Comparison of different architectures}\label{sec:mult-arch}
One of the advantages of the \texttt{\#pragma} based approach of OpenACC is that one code can be compiled for more than a single architecture. This section is dedicated to demonstrate the performance of \ac{SOMA} in different environments. \autoref{tab:arch} lists detailed information about the architectures used for the comparison. For our evaluation, we compiled the application with the highest available optimization flag of the corresponding platform. In all cases we used for this benchmark test a homopolymer melt comprised of $n=2^{14}$ chain molecules with $N=64$ particles each, resulting in a total number of $nN=1\,084\,576$ particles. In all cases we employed the parallelization over independent molecules. As shown in \autoref{fig:weak} we do not expect the ``SET'' algorithm to yield a significant improvement for these system sizes.

\begin{table}
  \centering
  \begin{tabular}{p{0.3\columnwidth}p{0.3\columnwidth}p{0.3\columnwidth}}
    \hline
    accelerator                & host \ac{CPU}               & compiler       \\
    \hline
    \hline
    \mbox{Nvidia Geforce} GTX480      & Intel Xeon E5620     & PGI 16.10 CUDA 8          \\
    \hline
    Nvidia Tesla K20           & AMD Opteron 6274            & PGI 16.10 CUDA 7.5 \\
    \hline
    \mbox{Nvidia Geforce} GTX580      & Intel Xeon E5620     & PGI 16.10 CUDA 8          \\
    \hline
    Intel Xeon Phi 7250        & --                          & Intel 2017          \\
    \hline
    multi-core \ac{CPU}        & 2 Intel Xeon E5-2680v3 & GCC 5.4           \\
    \hline
    Nvidia Tesla K80           & Intel Xeon E5-2609v4        & PGI 16.10 CUDA 8   \\
    \hline
    Nvidia Pascal P100         & Intel Xeon E5-2609   & PGI 16.10 CUDA 8          \\
    \hline
  \end{tabular}
  \caption{Details about the architectures used for comparison of the performance.}
  \label{tab:arch}
\end{table}

\begin{figure}
  \centering
  \includegraphics[width= \columnwidth]{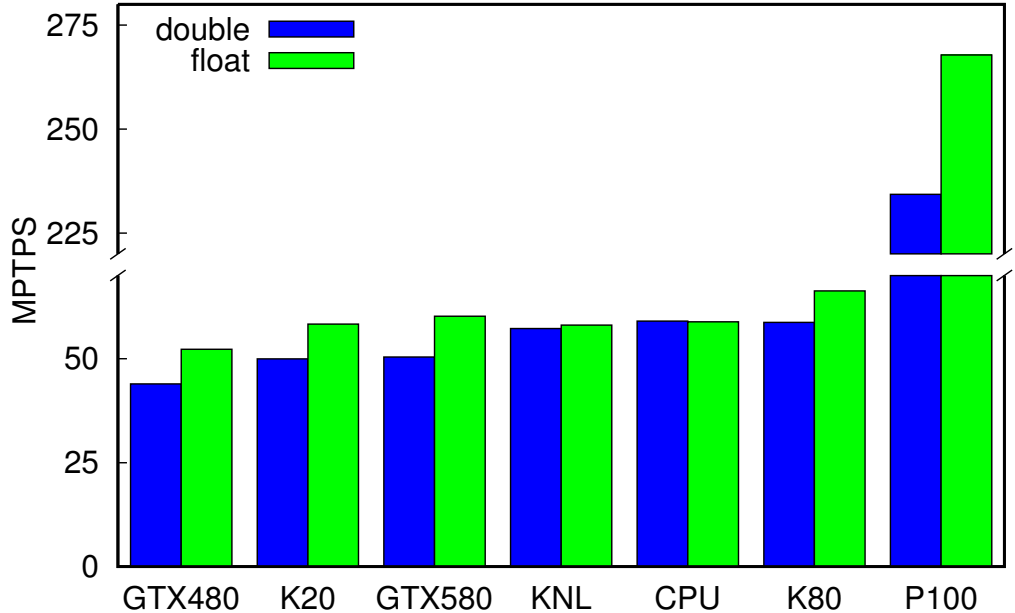}
  \caption{Performance comparison of different architectures.
    KNL is the abbreviation for the Intel Xeon Phi 7250 Knights Landing.
    CPU labels the multi-core performance of the Intel Xeon on a JURECA node.
    For more details about the architectures refer \autoref{tab:arch}.}
  \label{fig:arch}
\end{figure}

\autoref{fig:arch} plots the obtained performance in \ac{MPTPS}. Interestingly, the performance on the tested architectures is comparable, in spite of the fact that the release date of the architectures differs by more than five years. The exception is the Nvidia Pascal P100 accelerator, which roughly quadruples the performance compared to all other architectures. Although the code has not been optimized specifically for \acp{CPU} or the Knights Landing architecture, it is remarkable that the a simple dual socket Intel Xeon Haswell is slightly faster than the Intel Xeon Phi accelerator. We hypothesize that the memory throughput is the limiting factor for all simulations, explaining this characteristic.

\subsection{Combining Multiple Architectures}\label{sec:comb-mult-arch}
In the previous section, we have demonstrated that \ac{SOMA} can be accelerated by multiple types of available architectures. Furthermore, supercomputers often feature a heterogeneous configuration, i.e., a node comprises considerably more \ac{CPU} cores than accelerators. In this environment it is beneficial to assign to each accelerator a single \ac{MPI} rank with one associated \ac{CPU} core and, additionally, assign to all other cores on a node an additional \ac{MPI} rank.  For the example of a JURECA node with $24$ cores and $4$ accelerators this scheme assigns five \ac{MPI} ranks per node. The first four are each accelerated by a \ac{GPU} and the last combines the remaining $20$ \ac{CPU} cores using multiple threads.

As a proof of concept, we investigated a system of a single GK210 processor K80 accelerator in combination with a dual socket Intel Xeon E5-2680v3 with $24$ cores. The accelerator is managed with a single \ac{MPI} rank, and we gradually include in the second \ac{MPI} rank up to 23 \ac{CPU} threads. The executable for the accelerator is compiled with the PGI compiler and parallelized using OpenACC, whereas the multi-core \ac{CPU} executable is compiled with the GCC and parallelized using OpenMP.

\begin{figure}
  \centering
  \includegraphics[width=\columnwidth]{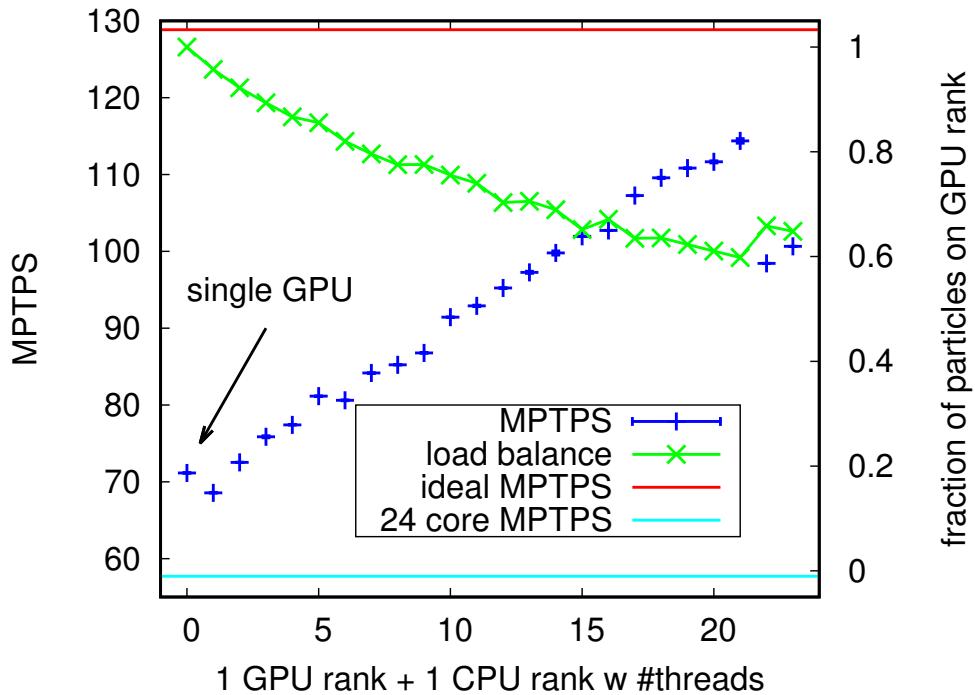}
  \caption{Combination of two heterogeneous \ac{MPI} ranks running a single simulation. The load balancer of \ac{SOMA} automatically distributes the load  to match performance of the two unequal
    ranks. For further details refer to\autoref{sec:comb-mult-arch}.
    The performance drop for 22 and 23 CPU cores can be explained with the NUMA architecture of the JURECA nodes.
    There is no idling core on each board available to coordinate communication between the two memories, thus the memory transfer is interfering with computation.}
  \label{fig:multi-arch}
\end{figure}

Not surprisingly, \autoref{fig:multi-arch} indicates that the performance drops if a second \ac{MPI} rank with only a single \ac{CPU} thread is coupled to the accelerated rank because of the increased communication and computational overhead. Utilizing already the second \ac{CPU} thread, however, the performance can be increased over that of the single accelerator. Thereafter, the performance scales linearly with the number of used \ac{CPU} threads. The total performance of bounded from below by the performance of the $24$ core \ac{CPU} and from above by the ideal sum of the single \ac{CPU} and single \ac{GPU} performance. Using the accelerator and the optimal number of \ac{CPU} cores, we obtain about $85\%$ of the ideal performance. This example demonstrates that using our multiple-architecture code is beneficial in heterogeneous environments.

In addition to the performance of this heterogeneous system, \autoref{fig:multi-arch} illustrates the capabilities of the automatic load balancer.
Because of the different computational powers of the two \ac{MPI} ranks, the load balancer transfers automatically polymer molecules from one rank to another, minimizing the synchronization time of the ranks. The linear dependence of fraction of molecules on a \ac{GPU} and the linear performance demonstrate that the load balancer, described in \autoref{sec:MPI-level}, works as expected.

\subsection{Comparison to Molecular Dynamics simulation }\label{sec:comp-molec-dynam}

To put the performance of \ac{SOMA} into some perspective, we compare its performance with another popular, publicly available \ac{MD} simulation package, HOOMD \cite{hoomd}. HOOMD has been designed for Nvidia \acp{GPU} by using CUDA, which makes it very efficient in accelerated environments. It is a general framework for \ac{MD} simulations of a broad variety of systems, including soft, coarse-grained model of multi-component polymer systems.

Whereas \ac{SOMA} uses soft, non-bonded, pairwise interactions that are evaluated on a collocation grid, see \autoref{eq:hamil-pair}, and the \ac{SCMF} algorithm explicitly exploits the time-scale separation between the strong, bonded and weak, non-bonded interactions (cf.~\autoref{sec:sampling-algorithm}), these two features are not available in \ac{MD} simulations using HOOMD. Therefore we utilize a different soft, coarse-grained model that represents a similar physical system, where the non-bonded interactions are represented by a soft, DPD-like, pairwise potential \cite{DPD2b,DPD1},
$V_\text{nb}(\boldsymbol{r})=\frac{k_\text{B}T}{2}v(1-|(\boldsymbol{r}|/\sigma)^2$ with $v=0.5$ that is cut-off at a distance $r\leq\sigma \approx 0.12 R_{e0}$.

For the simulations with \ac{SOMA} and HOOMD, we consider a homopolymer melt with chain discretization $N=128$ and invariant degree of polymerization $\sqrt{\bar{\cal N}} \approx 37$ and systematically vary the system size, $nN$. In order to quantify the performance of the \ac{MC} and \ac{MD} simulations,\footnote{Since the thermostat in a soft, coarse-grained model generates substantial friction and therefore slows the relaxation down, we do not employ a thermostat in the \ac{MD} simulations, i.e., the simulations are performed in the microcanonical ensemble.} we compare the computational execution time (in units of hours) it takes to relax the molecular conformations. The latter property can be estimated from the model time it takes for a molecule to diffuse its mean-squared end-to-end distance, $\langle R_{e}^{2}\rangle$. To this end, we monitor the mean-squared displacement, $g_{3}(t)$, of a molecule's center of mass and extract the self-diffusion coefficient, $D$, from the long-time behavior, $g_{3}(t)=6Dt$ for $t \to \infty$. The diffusion constant is measured for a single system because it is independent from the system size. We define the relaxation time in the soft, coarse-grained model as $t_{R}=\langle R_{e}^2\rangle / D$.

\begin{figure}
  \centering
  \includegraphics[width=\columnwidth]{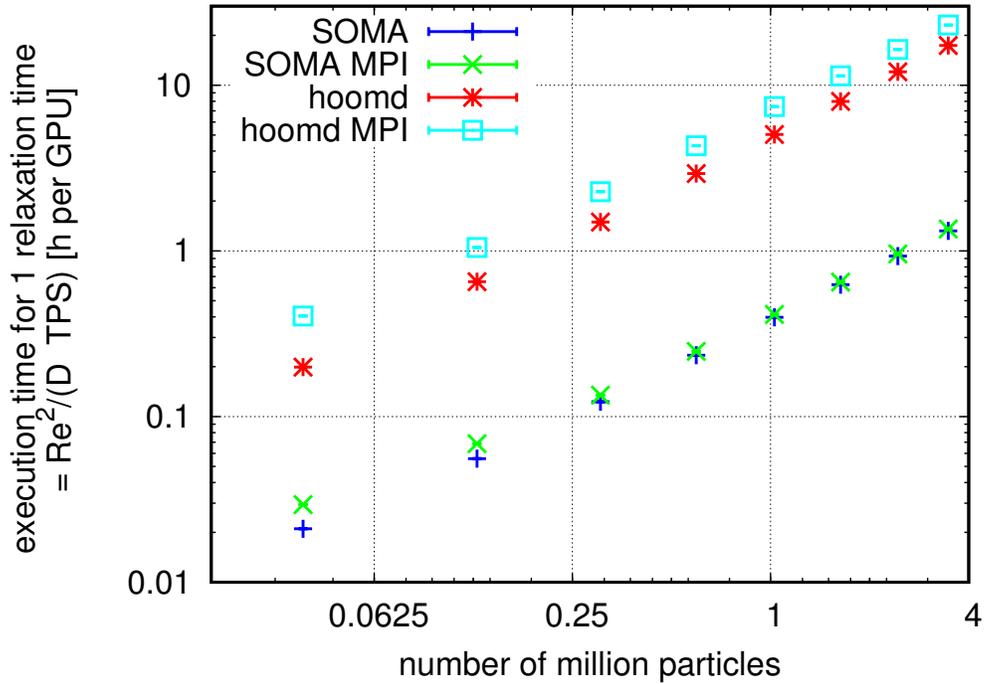}
  \caption{Comparison of the execution time required to simulate one relaxation time, $t_R$, of a homopolymer melt. The simulations are performed on one or two K80 \acp{GPU}, respectively. \ac{SOMA}
    simulations use a soft, coarse-grained model where the non-bonded interactions are evaluated on a collocation grid and the \ac{SCMF} algorithm is employed, whereas the HOOMD simulations use a
    DPD-model with soft interactions, representing a similar physical system. The data is scaled by the number GPU devices, highlighting that the multi-\acp{GPU} approach of \ac{SOMA} allowed by the
    \ac{SCMF} algorithm has a low overhead, compared to a domain decomposition.}
  \label{fig:hoomd}
\end{figure}

\autoref{fig:hoomd} shows the execution time required to relax the molecular conformations as a function of system size, $nN$, on K80 \ac{GPU}. Both programs exhibit a similar, good scaling, however, the execution of our soft, coarse-grained model using the \ac{SCMF} algorithm implemented in \ac{SOMA} is an order of magnitude faster than the \ac{MD} simulations using HOOMD. These results demonstrate the previously mentioned advantages of our soft, coarse-grained model and \ac{SCMF} algorithm. Additionally, \autoref{fig:hoomd} indicates that, for very small systems, the spatial domain decomposition scheme used for \ac{MPI}-based multi-\ac{GPU} simulations in HOOMD becomes inefficient, whereas the parallelization strategy of \ac{SOMA} still remains efficient.

\subsection{Application example: Self-assembly of diblock copolymers in thin films}
To wrap up our discussion we illustrate a prospective application of \ac{SOMA} to investigate the kinetics of self-assembly of diblock copolymers in response to a quench from the disordered state to
below the \ac{ODT}. In an experiment, such a process can be realized by a jump in temperature or solvent evaporation. As mentioned in the introduction, these flexible, linear molecules are comprised
of an $A$ and a $B$ block that repel each other. At sufficient thermodynamic incompatibility, below the \ac{ODT}, one observes microphase separation into spatially modulated phases \cite{Leibler80}. In the following
we consider two systems: (i) a lamella-forming, symmetric diblock copolymer with $A$-volume fraction, $f_A^\text{lam}=0.5$, and incompatibility, $\chi_{0}N^\text{lam}=17$, and (ii) a cylinder-forming
molecular architecture with $f_A^\text{hex}=0.75$, and $\chi_{0} N^\text{hex}=28$. In both systems, the contour of a molecule is discretized into $N=100$ coarse-grained particles, and the invariant
degree of polymerization takes the value $\sqrt{\mathcal{\bar{N}}} \approx 85.7$ for both systems. The relaxation time of a homopolymer in a comparable system is $t_R \approx
88.4\cdot 10^{3}$ \ac{MCS}. The relaxation time is determined as in \autoref{sec:comp-molec-dynam}.

The system is confined into a thin film with lateral dimensions, $L_x=L_y=200 R_{e0}$, and periodic boundary conditions are applied in the lateral $x$ and $y$ directions. In the third direction, $z$, the film is confined by two, planar, impenetrable and non-preferential surfaces that are spaced a distance $L_{z}$ apart. In order to stabilize standing, vertical structures, i.e., morphologies that do not significantly vary in $z$ directions, we chose the film thickness to be incompatible with a lying arrangement of lamellar sheets or cylinders -- $L_{z}^\text{lam}=1.2 R_{e0}$ and $L_z^\text{hex} = 0.75 R_{e0}$ for the lamella-forming and cylinder-forming copolymers, respectively. Thus, the lamellar and cylindrical systems comprise a total of $nN \approx 411 \cdot 10^6$ and  $240\cdot10^6$ particles. These systems contain many units cells of the spatially modulated phase, enabling the study of defect interaction and grain growth \cite{Alex16,Harrison2004, Harrison2000}.

The large necessary system size and particle number highlights the need of efficient simulation techniques, provided by \ac{SOMA}. While the morphologies that evolve after a quench from the disordered phase, $\chi_{0}N=0$, provide a wealth of information and can be analyzed in an automated fashion \cite{murphy2015automated}, a complete discussion is beyond the scope of the present manuscript, and we restrict ourselves to only highlighting some interesting characteristics.

\subsubsection{Lamella-forming system}\label{sec:lamell-conf}

\begin{figure}
  \centering
  \includegraphics[width=\columnwidth]{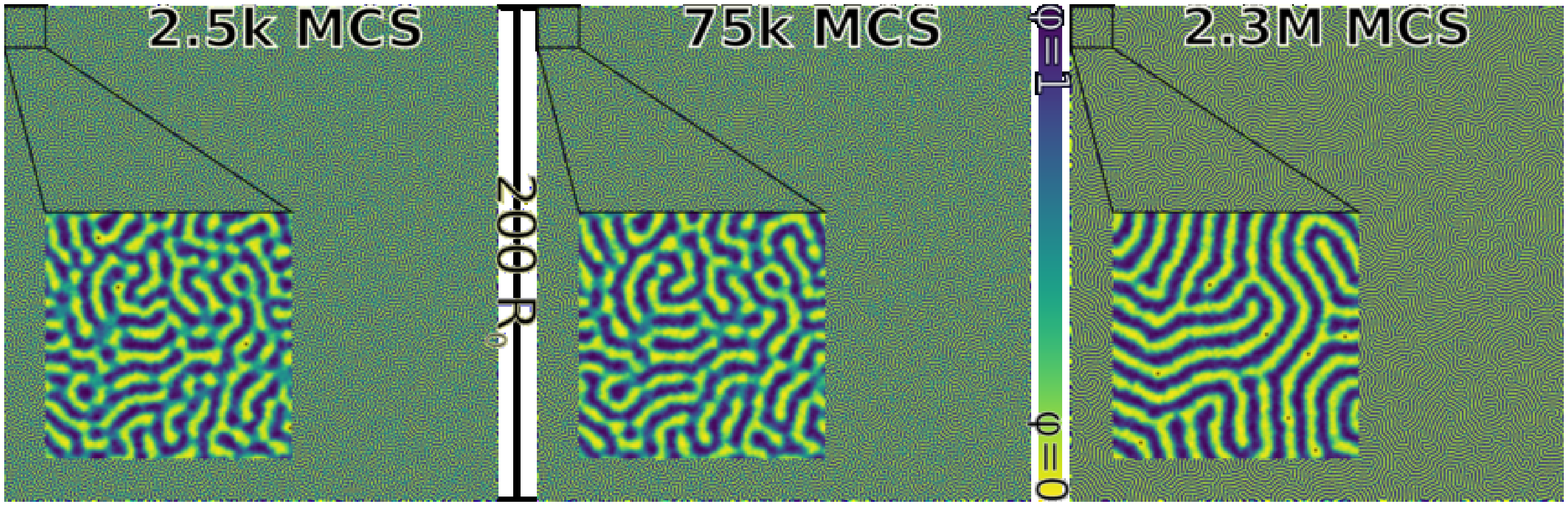}
  \caption{Time evolution of the composition, $\varphi(\boldsymbol{r}) = \frac{n_A(\boldsymbol{r})}{n_A(\boldsymbol{r})+n_B(\boldsymbol{r})}$, of in a  symmetric diblock copolymer thin film after a quench from the disordered state to $\chi_{0}N=17$. In the early stages, after $2\,500$ \ac{MCS}, a lamellar domains are formed. These domains form a fingerprint-like pattern that is riddled with defects ($7.5\cdot10^4$ \ac{MCS}). Further ordering, which proceeds via defect annihilation and grain growth ($2.3\cdot 10^6$ \ac{MCS}), is protracted. The enlarged insets highlight the local domain structure at the top, left corner.}
  \label{fig:lam-phi}
\end{figure}

The equilibrium configuration of symmetric block copolymers is the perfect lamellar state \cite{matsen2001standard}. The time evolution after a quench from the disordered state towards equilibrium at $\chi_{0} N =17$ is illustrated in \autoref{fig:lam-phi}, which depicts the spatially varying composition, $\phi(\boldsymbol{r})$. Immediately after the quench, domains form, in which $A$ or $B$ particles enrich. During this spinodal self-assembly, the local composition fluctuations exponentially increase in time until the composition inside the domains reaches its saturation values, $0$ or $1$. Whereas the local morphology consists of lines and stripes with a preferred distance, the correlation lengths is just a few lamellae. After $2\,500$ \ac{MCS}, the morphology is rather riddled with defects, and even grains with a well-defined orientation are difficult to identify. In the following, defects annihilate, the correlation length grows and a grain structure gradually emerges at $7.5\cdot10^4$ \ac{MCS}. Subsequently, the kinetics of structure evolution is extremely slow, i.e., the number of defects decreases only very gradually and the increase of grain size is protracted.

\begin{figure}
  \centering
  \includegraphics[width=\columnwidth]{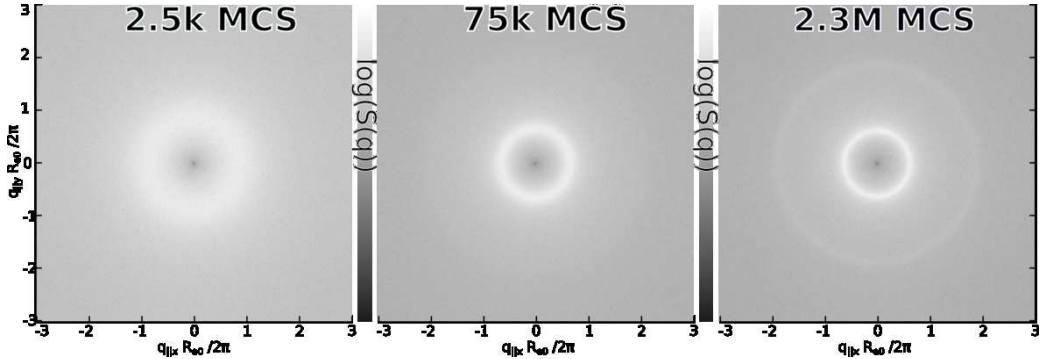}
  \caption{Time evolution of the two-dimensional structure factor $S(\boldsymbol{q}_{\|},t) \propto \left| \mathcal{F}[\phi_A - \phi_B] \right|^{2} $ after a quench of a symmetric diblock copolymer melt from the disordered phase. The image sequence matches the configurations in \autoref{fig:lam-phi}. The rings indicate the lamellar spacing $L_0$ and indicate that no long-range, preferential orientation of the lamellar domains has been established.}
  \label{fig:lam-sq}
\end{figure}

In experiments, the large-scale structure is often quantified by the time-dependent structure factor \cite{Michele17,Alex16},
\begin{align}
S(\boldsymbol{q}_{\|},t) & = \frac{N^2 \rho_0}{4 V } \left| \mathcal{F}[\phi_A - \phi_B] \right|^{2} \nonumber
\end{align}
which can be readily obtained from the Fourier transform, $\mathcal{F}$, of the composition field. The time evolution of the structure factor $S(\boldsymbol{q}_{\|},t)$ is presented in \autoref{fig:lam-sq} corresponding to the configurations of \autoref{fig:lam-phi}. Note that the system size is large enough to obtain a radially symmetric, two-dimensional structure factor from a single snapshot without averaging over different realizations of the stochastic time evolution. This clearly demonstrates that the finite system size does not influence the results.

At early times, the structure factor features a single, relatively broad ring indicating the characteristic length scale of the initial structure that results from the fastest growing mode of the spinodal structure formation. At the end of this spinodal self-assembly, the composition has saturated, and the ring in $S(\boldsymbol{q}_{\|},t)$ is indicative of a morphology with a characteristic length scale -- the distance between domains -- but no long-range order.  With the establishment of sharp interfaces between the lamellar domains, a second ring in $S(\boldsymbol{q}_{\|},t)$ becomes visible. Unfortunately, the time evolution is not long enough to establish a dominant, long-range orientation of the domains of the lamellar morphology; as a result, the rings stay uniform for all angles, $\theta$.

\begin{figure}
  \centering
  \includegraphics[width=\columnwidth]{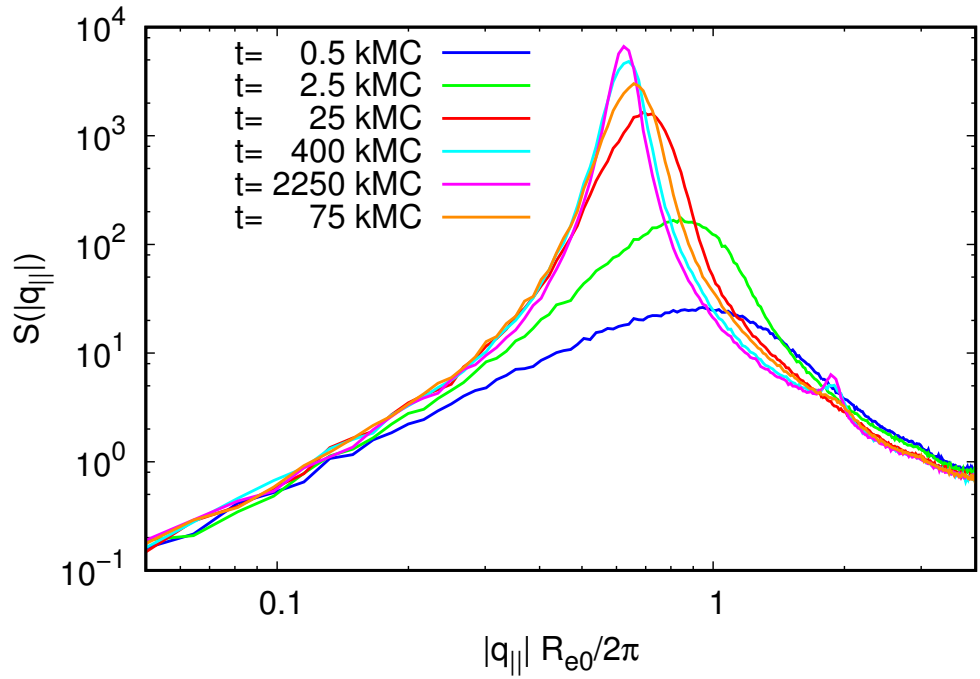}
  \caption{Radially averaged structure factor, $S(|\boldsymbol{q}_{\|}|,t)$, for the lamellar configuration. At early times the shift and narrowing of the dominant peak can be observed. At later times a second peak arises, signaling the sharpening of the internal domain interfaces.}
  \label{fig:sqr-lam}
\end{figure}

Since the large-scale structure is isotropic, the radially averaged structure factor, $S(|\boldsymbol{q}_{\|}|,t)$, is depicted in \autoref{fig:sqr-lam}. From the primary peak of $S(|\boldsymbol{q}_{\|}|,t)$ at $q_\text{max}$, we can extract the periodicity of the incipient lamellar structure, $\frac{L}{R_{e0}}=\frac{2\pi}{q_\text{max}R_{e0}}$.

We observe that $q_\text{max}$ decreases, i.e., the lamellar distance gradually increases with time. This behavior is expected because the fastest growing mode of the spinodal self-assembly, which dictates the distance between the incipient lamellae, occurs at a smaller wavevector than the wavevector that corresponds to the equilibrium lamellar spacing. Additionally, we observe that the primary peak in $S(\boldsymbol{q}_{\|},t)$ gradually sharpens, indicating the increase of the correlation length of the lamellar morphology. At late times, the primary peak does hardly evolve in time, i.e., the ordering process is protracted and the occasional defect motion and annihilation, e.g., merging of bridges between lamellae, do not result in the establishment of long-range order on the considered time scale.

The example illustrates the fascinating physics of ordering kinetics and clearly demonstrates the need of large system sizes -- in a system of smaller lateral extension defects would interact with themselves across the periodic boundary conditions via long-range strain fields and the growth of domains would be affected by finite-size effects. With the limited simulation time available, we are able to investigate the motion and annihilation of defects as well as the early stages of grain growth, as the insets of \autoref{fig:lam-phi} demonstrates, but we are unable to reach equilibrium, i.e., the lamellar structure with a vanishingly small equilibrium density of defects.

\subsubsection{Cylinder-forming system}\label{sec:hexag-cylind-phase}

\begin{figure}
  \centering
  \includegraphics[width=\columnwidth]{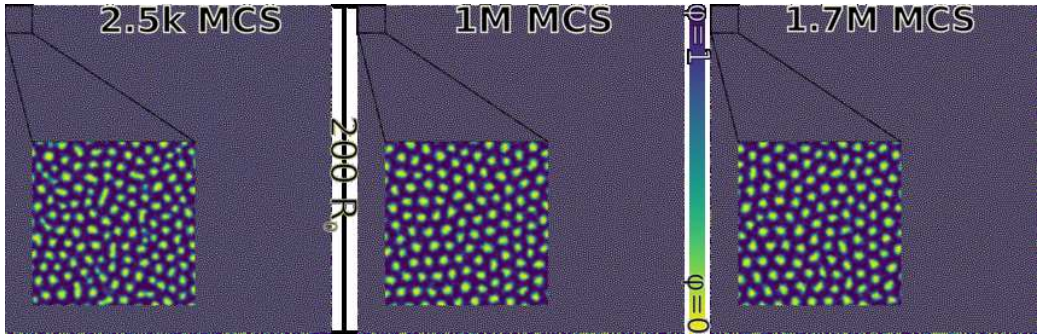}
  \caption{Time evolution of the order parameter $\varphi(\boldsymbol{r})$ of the hexagonal cylinder configuration. In the early stages, after $2\,500$ \ac{MCS}, cylindrical domains of the minority component form. Subsequently, grains of locally hexagonal orientation order emerge and the spatial position of the domains evolves as to optimize the hexagonal order. For the times accessible by our simulation, no long-range order is established.
  }
  \label{fig:hex-phi}
\end{figure}

The asymmetry of the volume fraction $f_A$ causes the polymer melt to form cylinders of the minority component. The judicious choice of the film thickness, $L_{z}$, forces the cylinders to stand upright in $z$ direction, i.e, top-down images of the film, presented in \autoref{fig:hex-phi}, provide direct insights into the kinetics of structure formation after a sudden change from the disordered phase to $\chi_{0} N^\text{hex}=28$.

The formation of the cylindrical domains can be observed in the early stages of time evolution. The system exhibits a fluid-like packing of domains, which locally resembles the hexagonal equilibrium structure \cite{matsen2001standard}, but no long-range order is established in the course of the simulation.

\begin{figure}
  \centering
  \includegraphics[width=\columnwidth]{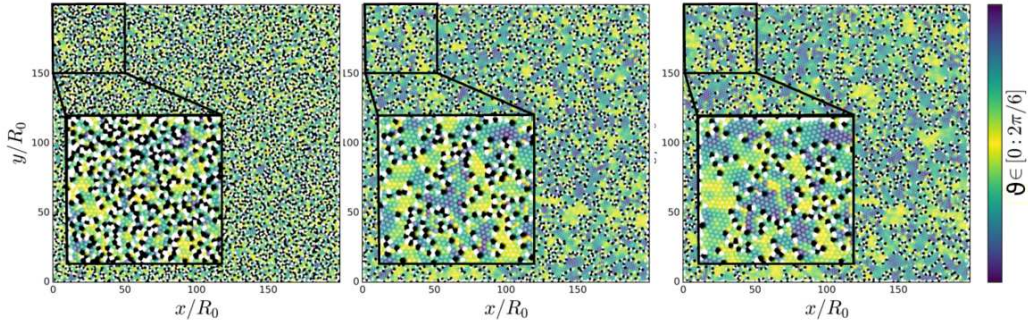}
  \caption{Voronoi tessellation of the cylindrical configuration. Hexagonal tiles are colored according to their orientation in the plane. Tiles with less than $6$ neighbors are colored white, whereas tiles with more the $6$ neighbors are colored black. These non hexagonal tiles indicate the interfaces between grains with different orientation. The analyzed configurations correspond to \autoref{fig:hex-phi}.
  }
  \label{fig:voronoi-large}
\end{figure}

The identification of hexagonal domains is not as straightforward as it is for lamellar structures. In order to visualize the grains of orientationally correlated cylinders, we make use of  Voronoi diagrams \cite{Li10,aurenhammer1991voronoi}. To this end, we tessellate the domain morphology so that each cylinder center is enclosed by straight lines, which separate its surrounding from its nearest neighbors. In a perfect, hexagonal lattice each cylinder has exactly $6$ neighbors, and we assign an orientation $\theta \in [0:\frac{2\pi}{6})$ in the $xy$-plane to each hexagon. \autoref{fig:voronoi-large} presents this Voronoi tessellation for the previously discussed configuration, where each hexagon is colored according to its orientation. Domains that do not have $6$ neighbors correspond to defects, and they are colored black or white. After the initial microphase separation, grains of hexagonal domains with correlated orientation become visible, and the size of these grains grows in time. Non-hexagonal domains are preferentially located at grain boundaries, and their number decreases in the course of coarsening.

\begin{figure}
  \centering
  \includegraphics[width=\columnwidth]{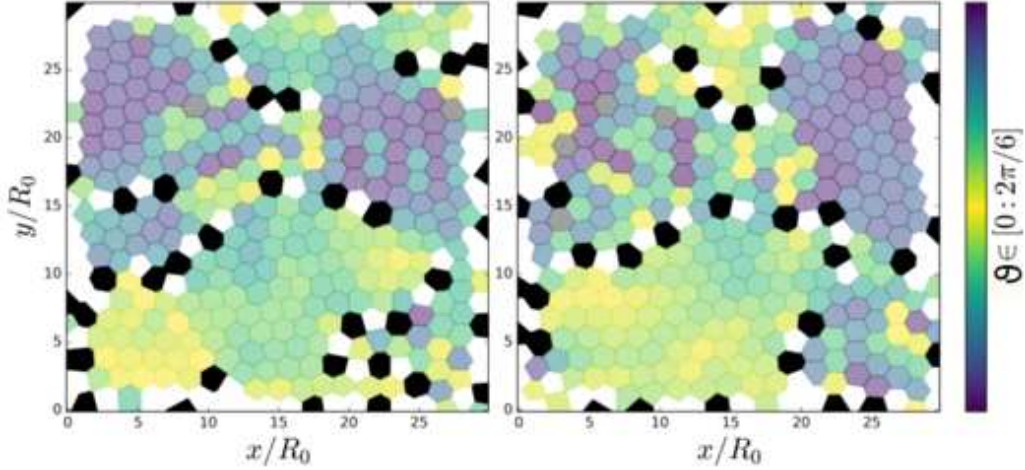}
  \caption{Voronoi tessellation of a smaller ($30 R_{e0}\times 30 R_{e0} \times 0.75 R_{e0}$), but otherwise identical system of cylinder-forming copolymers. The left configuration is a snapshot after $1.5\cdot 10^6$ \ac{MCS} after the quench from the disordered state, while the right plot corresponds to $5\cdot10^6$ \ac{MCS}. The longer simulation times highlights the protracted coarsening of the grains after their establishment.}
  \label{fig:small-hex}
\end{figure}

For further insights to the time evolution of the domain in the hexagonal configuration, we investigated a smaller configuration $30 R_{e0}\times 30 R_{e0} \times 0.75 R_{e0}$ with $nN = 5.4\cdot10^{6}$ particles. We simulated this configuration for $5 \cdot 10^{6}$ \ac{MCS}, covering a much longer period than the previous two figures. \autoref{fig:small-hex} depicts the Voronoi tessellation for the smaller system, where we observe the slow coarsening of grains.  The comparison of the two panels at $1.5  \cdot 10^{6}$ \ac{MCS} and $5 \cdot 10^{6}$ \ac{MCS} demonstrates that the grains remain almost identical, but smaller defects in the boundaries between grains have been expunged.

\begin{figure}
  \centering
  \includegraphics[width=\columnwidth]{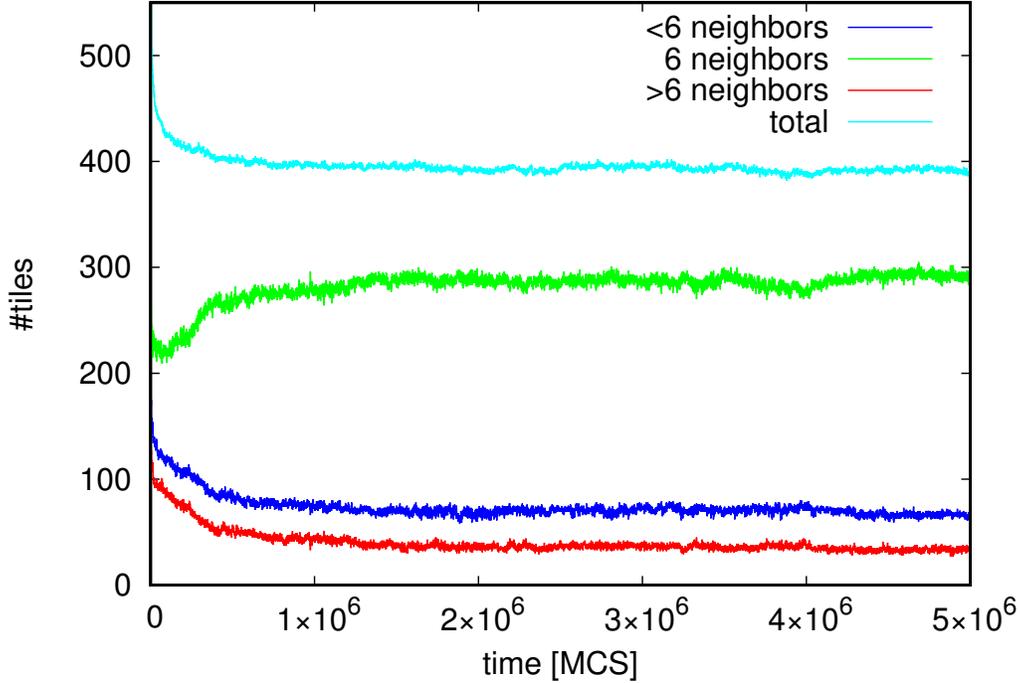}
  \caption{Time evolution of perfect and imperfect tiles of the Voronoi tessellation. It is not expected that the number of imperfect tiles approaches zero for the perfect hexagonal lattice, because even though the film is periodic the analysis is not; introducing artifacts at the boundaries.}
  \label{fig:small-tiles}
\end{figure}

Using the Voronoi analysis we quantify the degree of hexagonal order by the number of domains or tiles, whose number of neighbors differs from $6$. \autoref{fig:small-tiles} plots the number of tiles with less than $6$ neighbors, $6$ neighbors and more than $6$ neighbors as a function of time. At very short times the total number of cylindrical domains rapidly decreases and a the near-equilibrium domain size of the cylindrical phase is established. Subsequently, on an intermediate time scale, the number of perfect tiles with $6$ neighbors increases and the number of imperfect tiles decreases, respectively. On this time scale, the system is forming grains, and the reduction of the less or more than $6$-fold coordinated domains indicates that the grain boundary length decreases, i.e., the grain size grows. After $1.5  \cdot 10^{6}$ \ac{MCS} the number of $6$-fold hexagons hardly increases indicating that the remaining dynamics is almost arrested.

Although, we focused our analysis in the cylindrical and lamellar example only on the spatial compositions, the particle based model enables the investigation of the chain configurations at all times.
Especially, a connection between the single chain dynamics and the collective ordering of the system can be made without the need of an Onsager coefficient.

\section{Conclusions}\label{sec:conclusion}

We have presented an efficient implementation of the \ac{SCMF} algorithm for the simulation of soft, coarse-grained polymer models, which scales well for both, modern \ac{CPU}-based cluster architectures as well as clusters based on Nvidia \ac{GPU} accelerators. Using the OpenACC model in conjunction with \ac{MPI} allows us to use a variety of different compute environments.

The software allows us to study large systems with billions of particles and thereby enables researches to investigate scientific questions in the wide area of soft-matter and self-assembly using state-of-the-art supercomputers. The software is available under the GNU Lesser General Public License version 3. We are planning to further develop \ac{SOMA}, to integrate additional features, and to tune execution efficiency for other accelerators and invite researchers at other institutions to use the program and contribute to its development.

\section{Acknowledgment}\label{sec:acknowledgement}
We thank the all co-developers of \ac{SOMA}, especially Marcel Langenberg, Fabien L\'eonforte, Juan Carlos Orozco Rey, and Ulrich Welling as the other main developers for their contributions. Additionally, we acknowledge useful and instructive discussion on \ac{GPU} optimizations of the code with St\'ephane Chauveau and Brent Lebeck. We also thank the \ac{ORNL}, the Technische Universit\"at Dresden and the \ac{JSC} for organizing the Hackathon 2016, at which we started porting this program to accelerators using OpenACC.

The authors gratefully acknowledge the computing time granted by the John von Neumann Institute for Computing (NIC) and provided on the supercomputer JURECA \cite{krause2016jureca} at \ac{JSC}. Additionally, this research used resources of the Oak Ridge Leadership Computing Facility at the Oak Ridge National Laboratory, which is supported by the Office of Science of the U.S. Department of Energy under Contract No.DE-AC05-00OR22725. Financial support has been provided by \ac{DFG} under grant Mu 1674/16-1.


\begin{acronym}[MPTPS]
  \acro{CPU}{Central Processing Unit}
  \acro{CUDA}{Compute Unified Device Architecture}
  \acro{GPGPU}{General Purpose Computing on Graphics Processing Units}
  \acro{MSD}{Mean-Squared Displacement}
  \acro{PRNG}{Pseudo Random Number Generation}
  \acro{ALU}{Arithmetic Logic Unit}
  \acro{DPD}{Dissipative Particle Dynamics}
  \acro{GPU}{Graphics Processing Unit}
  \acro{MC}{Monte-Carlo}
  \acro{MD}{Molecular Dynamics}
  \acro{MFA}{Mean-Field Approximation}
  \acro{MPTPS}{Million Particle Timesteps Per Second}
  \acro{MPI}{Message Passing Interface}
  \acro{PCG}{Permuted Congruential Generator}
  \acro{MT}{Mersenne-Twister}
  \acro{SCMF}{Single-Chain-in-Mean-Field}
  \acro{SCFT}{Self-Consistent Field Theory}
  \acro{SMC}{Smart Monte-Carlo}
  \acro{JSC}{J\"ulich Supercomputing Centre}
  \acro{DFG}{Deutsche Forschungs Gemeinschaft}
  \acro{ORNL}{Oak Ridge National Laboratory}
  \acro{HDF5}{Hierarchical Data Format version 5}
  \acro{IO}{Input/Output}
  \acro{XDMF}{Extensible Data Model and Format}
  \acro{XML}{Extensible Markup Language}
  \acro{DSA}{Directed Self-Assembly}
  \acro{HPC}{High Performance Computing}
  \acro{SOMA}{SOft coarse grained Monte-carlo Acceleration}
  \acro{ODT}{Order-Disorder-Transition}
  \acro{MCS}{Monte-Carlo steps}
  \acro{BCC}{Body-Centered Cubic}
\end{acronym}
\end{document}